\documentstyle[aps,preprint,prl]{revtex}
\begin{document}
\flushbottom

\widetext
\draft
\title{Exact Dirac equation calculation of ionization 
induced by ultrarelativistic heavy ions}
\author{A. J. Baltz}
\address{
Physics Department,
Brookhaven National Laboratory,
Upton, New York 11973}
\date{\today}
\maketitle

\def\thepage{\arabic{page}}
\makeatletter
\global\@specialpagefalse
\ifnum\c@page=1
\def\@oddhead{Draft\hfill To be submitted to Phys. Rev. C}
\else
\def\@oddhead{\hfill}
\fi
\let\@evenhead\@oddhead
\def\@oddfoot{\reset@font\rm\hfill \thepage \hfill}
\let\@evenfoot\@oddfoot
\makeatother

\begin{abstract}
The time-dependent Dirac equation can be solved exactly for ionization
induced by ultrarelativistic heavy ion collisions.  Ionization calculations
are carried out in such a framework for a number of representative ion-ion
pairs.  For each ion-ion pair, the computed cross section consists of two
terms, a constant energy independent term and a term whose coefficient is
$\ln \gamma$.  Scaled values of both terms are found to decrease with
increasing Z of the nucleus that is ionized.
\end{abstract}

\makeatletter
\global\@specialpagefalse
\def\@oddhead{\hfill}
\let\@evenhead\@oddhead
\makeatother
\nopagebreak
\section{Introduction}
In a recent work\cite{ajb1} an exact semiclassical Dirac equation calculation
of ionization probabilities was presented.  The case considered was 
colliding Pb + Pb at ultrarelativistic energies.  A single electron was
taken to be bound to one nucleus with the other nucleus completely
stripped.  The probability that the electron would be ionized in the
collision was calculated as a function of impact
parameter, but no cross sections were presented.  In this paper the approach
of Ref. \cite{ajb1} is extended in order to calculate ionization cross
sections for a number of representative cases of collisions involving
Pb, Zr, Ca, Ne and H ions.  In Section II the exact semiclassical method
is reviewed and calculations of impact parameter dependent ionization
probabilities are presented.  In Section III the results of the probability
calculations are used to construct cross sections for various ion-ion
collision combinations in the form
\begin{equation}
\sigma =  A \ln \gamma + B
\end{equation}
where $A$ and $B$ are constants for a given ion-ion pair and
$\gamma\ (=1/\sqrt{1-v^2})$ is the
relativistic factor one of the ions seen from the rest frame of the other.
Comparisons are made with previous ionization calculations.
In Section IV the example of Pb + Pb at the AGS, CERN SPS,
and RHIC is worked out and the CERN SPS case is compared with data.

\section{Impact parameter dependent probabilities}
If one works in the appropriate gauge \cite{brw}, then
the Coulomb potential produced by an ultrarelativistic particle (such as a
heavy ion) in uniform motion can be expressed in the following form\cite{ajb}
\begin{equation}
V(\mbox{\boldmath $ \rho$},z,t)
=-\delta(z - t) \alpha Z_P(1-\alpha_z)
\ln{({\bf b}-\mbox{\boldmath $ \rho$})^2 \over b^2 }.
\end{equation}
${\bf b}$ is the impact parameter, perpendicular to the $z$--axis along which
the ion travels, $\mbox{\boldmath $\rho$}$, $z$, and $t$ are the coordinates of
the potential relative to a fixed target (or ion),
$\alpha_z$ is the Dirac matrix, $\alpha$ is the fine structure constant,
with $Z_P$ and $v$ the charge and velocity of the moving ion.
This is the physically relevant ultrarelativistic potential since it was 
obtained by ignoring terms in
$({\bf b} - \mbox{\boldmath $\rho$}) / \gamma^2$\cite{ajb}\cite{brw}.  As will
be shown in Section II, when
${\bf b}$ becomes large enough that the expression Eq.(2) is inaccurate, we
match onto a Weizsacker-Williams expression which is valid for large
$b$.  Note that the $b^2$ in the denominator of the logarithm in Eq.(2) is
removable by a gauge transformation, and we retain the option of keeping or
removing it as convenient.

It was shown in Ref. \cite{ajb1} that the $\delta$ function allows the
Dirac equation to be solved exactly at the point of interaction, $z = t$.
Exact amplitudes then take the form
\begin{eqnarray}
a_f^j(t=\infty)&=\delta_{fj} & +
\int_{-\infty}^{\infty} d t e^{i (E_f - E_j) t} \langle \phi_f \vert
\delta(z - t) (1-\alpha_z) \nonumber \\ 
&&\times ( e^{-i \alpha Z_P \ln{({\bf b}- \mbox{\boldmath $ \rho$})^2 }} - 1 ) 
\vert \phi_j \rangle
\end{eqnarray}
where $j$ is the initial state and $f$ the final state.  This ampltude is in
the same form as the perturbation theory amplitude, but with
an effective potential to represent all the higher order effects exactly,
\begin{equation}
V(\mbox{\boldmath $ \rho$},z,t)
=-i \delta(z - t) (1-\alpha_z)
( e^{-i \alpha Z_P \ln{({\bf b}- \mbox{\boldmath $ \rho$})^2 }} - 1 ) ,
\end{equation}
in place of the potential of Eq.(2).  

Since an exact solution must be unitary,
the ionization probability (the sum of probabilities of excitation from the
single bound electron to particular continuum states) is equal
to the deficit of the final bound state electron population
\begin{equation}
\sum_{ion} P(b) = 1 - \sum_{bound} P(b)
\end{equation}
The sum of bound state probabilities includes the probability that the electron
remains in the ground state plus the sum of probabilities that it ends up in an
excited bound state.
From
Eq.(3) one may obtain in simple form the exact survival probability of an
initial state
\begin{equation}
P_j (b) =  \vert \langle \phi_j \vert (1-\alpha_z)
e^{-i \alpha Z_P \ln{({\bf b}- \mbox{\boldmath $ \rho$})^2 }} 
\vert \phi_j \rangle \vert^2.
\end{equation}
In principle the ionization probability can be calculated without
reference to final continuum states.  In practice ionization will be 
calculated both as a sum of continuum probabilities as well as the deficit
of bound state probabilities.

Table I shows the results of ionization calculations for an ultrarelativistic
Pb + Pb collision.  One of the Pb ions has an electron initially in the ground
state.  The other is completely ionized.   Final state probabilities for the
electron are calculated as a function of impact parameter $b$. 
Calculations have been carried out with a logarithmic
spacing in values of $b$, with the range of $b$ chosen to go from constant
probabiity of ionization at the low end to $1/b^2$ behavior at the high end.
The last
column which is the sum of final bound state (column 3) and final continuum
state (column 4) population exhibits a small
deficit from unity, presumably mostly from the truncation of the energy sum
over excited continuum states or secondarily from the truncation of the
bound state sum in the calculations.

Tables II and III show corresponding calculations for Ca + Ca, and Ne + Ne.  

\section{Cross sections}
The actual cross section comes from the impact parameter integral
\begin{equation}
\sigma_{ion}  = 2 \pi \int P(b)\ b\ d b. 
\end{equation}
As was exemplified in Tables I-III,
for each ion-ion case calculation of probabilities was performed at ten
impact parameters, in a scheme of sequential doubling.  The points
ran from an impact parameter
small enough that the probabilities were approximately constant with $b$, to
an impact parameter large enough that the probabilities had started to fall
off as $1 / b^2$.  The part of the integral, Eq.(7), over this region from the
first to the tenth impact parameter takes the form of sum of nine integrals
on a logarithmic scale  
\begin{eqnarray}
\sigma_{1-10} &=& \sum_{i=1,9} 2 \pi \int_{b_i}^{b_{i+1}} P(b)\ b\ d b 
\nonumber \\ 
&\simeq& \sum_{i=1,9} 2 \pi < P(b) b^2 > \int_{b_i}^{b_{i+1}} {d b \over b}. 
\end{eqnarray}
Approximating $< P(b) b^2 >$ over each interval by the average of the end
points we have (for $b_{i+1} = 2 b_i$)
\begin{equation}
\sigma_{1-10} = \pi \ln{2} \sum_{i=1,9} (P(b_i) b_i^2 + P(b_{i+1}) b_{i+1}^2). 
\end{equation}

Since the probability goes to a constant at the lowest impact parameter, $b_1$,
the contribution to Eq. (7) from zero up to $b_1$ is given simply by
\begin{equation}
\sigma_{0-1} = \pi P(b_1) b_1^2.
\end{equation}

We now need the contribution from the last point computed, $b_{10}$, out to
where $P(b)$ cuts off.  We match a Weizsacker-Williams type calculation to the
exact calculation at this $b_{10}$ in order to determine the
normalization for the calculation of probabilities at larger impact
parameters and to make the high end cutoff in $b$.  How the calculation in
the delta function gauge and the
calculation in the Weizsacker-Williams formulation are equivalent at this
matching impact parameter $b_{10}$ is presented in Appendix A.

The Weizsacker-Williams expression for a transition probability
at a given impact parameter is of the form
\begin{equation}
P_{WW}(b) = \int_{E_B}^\infty P(\omega) {\omega^2 \over \gamma^2}
K_1^2({\omega b \over \gamma}) d \omega,
\end{equation}
with $E_B$ the ground state electron binding energy
(a positive number here).

If $b \omega << \gamma$ then
\begin{equation}
K_1^2({\omega b \over \gamma}) = {\gamma^2 \over \omega^2 b^2},
\end{equation}
and
\begin{equation}
P_{WW}(b) = {1 \over b^2} \int_{E_B}^\infty P(\omega) d \omega.
\end{equation}
At the matching impact parameter, Eqns. (12) and (13) are accurate up to the
point where the
energy starts to cut off.  Thus $\int_{E_B}^\infty P(\omega) d \omega$ may
be simply determined
\begin{equation}
\int_{E_B}^\infty P(\omega) d \omega = b_{10}^2 P(b_{10}).
\end{equation}
Next recall that to high degree of accuracy
\begin{equation}
{\omega^2 \over \gamma^2} \int_{b_{10}}^\infty K_1^2({\omega b \over \gamma})
b\ db = ln({.681 \gamma \over \omega b_{10}}).
\end{equation}
Then from Eqns. (7), (11), (14), and (15)
the contribution to the cross section for impact parameters greater than
$b_{10}$ is
\begin{eqnarray}
\sigma_{10-\infty}&=& 2 \pi \Biggl( \ln({.681 \gamma \over b_{10}})
\int_{E_B}^\infty P(\omega) d \omega - \int_{E_B}^\infty P(\omega) \ln{\omega}
d \omega \Biggr)\nonumber \\
&=& 2 \pi b_{10}^2 P(b_{10}) \Biggl(\ln({.681 \gamma \over b_{10}}) 
- <\ln{\omega}>\Biggr).
\end{eqnarray}
$<\ln\omega>$ can be evaluated from the empirical observation that
at $b_{10}$, $P(\omega)$ goes as $1 / \omega^n$ with $n \simeq 3.8$.
One obtains
\begin{equation}
<\ln{\omega}> = \ln{E_B} + {1 \over (n - 1)}
\end{equation}
One now has the full ionization cross section
\begin{equation}
\sigma = \sigma_{0 - 1} + \sigma_{1-10} + \sigma_{10-\infty}
\end{equation}
or in the usual form
\begin{equation}
\sigma = A\ \ln{\gamma} + B.
\end{equation}
with
\begin{equation}
A = 2 \pi b_{10}^2 P(b_{10}).
\end{equation}
and
\begin{eqnarray}
B& =& \pi P(b_1) b_1^2 + \pi \ln{2} \sum_{i=1,9} (P(b_i) b_i^2 
+ P(b_{i+1}) b_{i+1}^2) \nonumber \\
&& + 2 \pi  P(b_{10}) b_{10}^2 \Biggl (\ln({.681 \over b_{10}}) 
- \ln{E_B} - {1 \over (n - 1)}\Biggr). 
\end{eqnarray}

The $A \ln{\gamma}$ term is entirely from the non-perturbative, large impact
parameter region and gives the beam energy dependence arising from the impact
parameter cutoff at $b \simeq \gamma / \omega$ (see Eqns. (15) and (16)).
Despite the form of Eq.(20) $A$ does not really depend on the matching impact
parameter $b_{10}$ since $b_{10}$ is in the region where $P(b) \sim 1 /
b^2$.  The $B$ term is independent of beam energy and contains
non-perturbative components from the smaller impact parameters.

Table IV shows the results of calculations of the ionization cross section
components $A$ and $B$ for symmetric ion-ion pairs.  There is good agreement
between the cross sections calculated by subtracting the bound state
probabilities from unity (first rows) or calculated by summing continuum
electron final states (second rows).  The agreement with the Anholt and Becker
calculations\cite{ab} in the literature is good for the lighter species for
both $A$ and $B$.
However with increasing mass of the ions the perturbative energy dependent
term $A$ decreases in the present calculations whereas it increases in the
Anholt and Becker calculations.  The greatest discrepancy is for Pb + Pb,
with Anholt and Becker being about 60\% higher.  Perhaps this discrepancy
is due to the fact that Anholt and Becker use approximate relativistic bound
state wave functions and the present calculations utilize exact
Dirac wave functions for the bound states.  Surprisingly, it is the term
$B$ (which has the non-perturbative component) where agreement is relatively
good between Anholt and Becker and the present calculations of Table IV.

Table V shows results of the calculation of $B$ (multiplied by $Z_2^2/Z_1^2$)
for a number of representative
non-symmetric ion-ion pairs.  (Since $A$ is perturbative, scaling as
$Z_1^2$, its value can be taken from Table IV for the various pairs here.)
Note that if one goes to the perturbative limit for a Pb target to be ionized
(H + Pb or Ne +Pb) then the scaled $B$ values (17,090, 17,030) are
some 30\% higher than the necessarily perturbative Anholt and Becker value of
13,000.  The good agreement of Anholt and Becker with the present calculations
for Pb + Pb $B$ seen in Table IV is thus somewhat fortuitous.

\section{An example}
Table VI presents calculated ionization cross sections for Pb + Pb at the
AGS, the CERN SPS, and RHIC.  CERN SPS data of Pb with a single electron
impinging on a Au target has recently been
published by Krause et al.\cite{kr}.  
Their measured cross section of 42,000 barns is
significantly smaller than the Anholt and Becker calculation (which includes
screening in the target Au) of about 63,600 barns.  
The result of the present Pb + Pb
calculation (55,800 to 58,200 barns) does not include screening and should
be compared with the corresponding no-screening calculation of Anholt and
Becker (83,700 barns).  What was essentially the present Pb + Pb result was
privately
communicated to Krause et al., and they seem to have then assumed that if
screening
were to be included, then the present calculation should be scaled by
the ratio of the Anholt and Becker screened to unscreened results.
They comment in their paper,
``With screening included\cite{ab} and scaled to a Au target, the Baltz value
agrees with the $\sigma_i$ measured in the ionization experiment
($4.2 \times 10^4$ b).''

\section{Acknowledgments}
I would like to thank Raju Venugopalan for several comments that helped clarify
the presentation of this work. 
This manuscript has been authored under Contract No. DE-AC02-98CH10886 with
the U. S. Department of Energy. 
\vfill\eject
\appendix
\section{Relationship of the $\delta$ function potential to the
Weizsacker-Williams formulation}
The Weizsacker-Williams cross section for a process induced by a heavy ion
projectile of charge $Z_p, \sigma_{ww}(\omega)$ is expressed in terms of the
photoelectric cross section $\sigma_{ph}(\omega)$ for the same process.
\begin{equation}
\sigma_{ww}(\omega) = 2 \pi \int_{b_0}^\infty {\alpha Z_p^2 \over \pi^2}
{\omega \over \gamma^2} \sigma_{ph}(\omega) K_1^2({b\,\omega \over  \gamma})
\ b \ d b
\end{equation}
Now since the photoelectric cross section for a process is given by
\begin{equation}
\sigma_{ph}(\omega) = {4 \pi^2 \alpha \over \omega} \vert \int d^3 r \psi^*_f
\mbox{\boldmath $ \alpha$} \cdot \hat{{\bf e}} \psi_0 e^{i \omega z} \vert^2,
\end{equation}
the Weizsacker-Williams amplitude (apart from an arbitrary constant phase) for
the process is
\begin{equation}
a_{ww}  = { 2 \alpha Z_p \over \gamma} 
K_1({b\,\omega \over  \gamma}) \int d^3 r \psi^*_f
\mbox{\boldmath $ \alpha$} \cdot \hat{{\bf b}} \psi_0 e^{i \omega z}
\end{equation}

Consider the $\delta$ function gauge
\begin{equation}
V(\mbox{\boldmath $ \rho$},z,t)
=-\delta (z - t) \alpha Z(1-\alpha_z)
\ln{({\bf b}-\mbox{\boldmath $ \rho$})^2 \over b^2}.
\end{equation}
Its multipole expansion is
\begin{eqnarray}
V(\mbox{\boldmath $ \rho$},z,t)&=&\alpha Z(1-\alpha_z) 
\delta (z - t) \nonumber \\
&& \biggl\{ -\ln {\rho^2 \over b^2}\ \ \ \ \ \ \ \ \ \ \ \ \  \rho>b\nonumber\\
&&+\sum_{m>0}{2 \cos m \phi \over m}\nonumber\\
&&\times\biggl[\biggl({\rho \over b}\biggr)^m
 \ \ \ \ \ \rho <b \nonumber\\
&&+\biggl({b \over \rho}\biggr)^m\biggr]\biggr\}.
 \ \ \ \ \ \rho >b
\end{eqnarray}
For $ b >> \rho $
\begin{equation}
V(\mbox{\boldmath $ \rho$},z,t)
=\delta (z - t) \alpha Z(1-\alpha_z) 
2 {\rho \over b} \cos \phi .
\end{equation}

One may make a gauge transformation on the wave function 
\begin{equation}
\psi=e^{-i\chi({\bf r},t)} \psi'
\end{equation}
where
\begin{equation}
\chi({\bf r},t)
=-2 \theta(t - z) \alpha Z {\rho \over b} \cos \phi .
\end{equation}
This leads to added gauge terms in the transformed potential
\begin{equation}
- {\partial \chi({\bf r},t) \over \partial t }
- \mbox{\boldmath $\alpha$} \cdot {\bf \nabla} \chi({\bf r},t)
=2 \delta(z - t) (1-\alpha_z)  \alpha Z {\rho \over b} \cos \phi
+ 2 \theta(t - z) \alpha Z { \mbox{\boldmath $\alpha$} \cdot \hat{{\bf b}}
\over b}.
\end{equation}
(This is the same transformation as previously carried out without the
restriction $ b >> \rho$ to go to the light cone gauge\cite{bm}.)
Here we obtain the light cone gauge potential for $ b >> \rho $
\begin{equation}
V(\mbox{\boldmath $ \rho$},z,t) = 2 \theta(t - z) \alpha Z
{ \mbox{\boldmath $\alpha$} \cdot \hat{{\bf b}}\over b},
\end{equation}
and we then obtain the perturbative amplitude in the light cone gauge
\begin{equation}
a_{cone}  = {- 2 i \alpha Z_p \over b} \int_z^{\infty} d t \int d^3 r \psi^*_f
\mbox{\boldmath $ \alpha$} \cdot \hat{{\bf b}} \psi_0 e^{i \omega t}.
\end{equation}
Integrate over $t$
\begin{equation}
a_{cone}  = { 2 \alpha Z_p \over \omega b} \int d^3 r
\psi^*_f \mbox{\boldmath $ \alpha$} \cdot \hat{{\bf b}} \psi_0 e^{i \omega z}.
\end{equation}

Now consider $a_{ww}$.  For $ \gamma >> b \omega $ 
\begin{equation}
K_1({b\,\omega \over  \gamma}) = { \gamma \over b\,\omega },
\end{equation}
and Eq. (A3) becomes
\begin{equation}
a_{ww}  = { 2 \alpha Z_p \over \omega b} \int d^3 r
\psi^*_f \mbox{\boldmath $ \alpha$} \cdot \hat{{\bf b}} \psi_0 e^{i \omega z}.
\end{equation}

Thus if one transforms from the delta function gauge to the light cone gauge
the amplitude in that light cone gauge is found to be equal to the
Weizsacker-Williams amplitude (within an arbitrary 
constant phase) as long as $ b >> \rho $ and $ \gamma >> b \omega $.

\begin{table}
\caption[Table I]{Ionization and Unitarity: Probabilities for Pb + Pb}
\begin{tabular}{ccccc}
b(fm) & $e^-_{gr}$ & $\sum_{bnd} e^-$ & 
$\sum_{cont} e^-$ &
$\sum e^- $\\
62.5 & .4344 & .5337 & .4399& .9736 \\
125 & .4467 & .5474 & .4228 & .9703 \\
250 & .4884 & .5920 & .3788 & .9708 \\
500 & .5820 & .6828 & .2907 & .9735 \\
1000 & .7303 & .8165 & .1691 & .9856 \\
2000 & .8899 & .9447 & .0526 & .9973 \\
4000 & .97056 & .98986 & .00987 & .99972 \\
8000 & .99270 & .99777 & .00217 & .99994 \\
16,000 & .998178 & .999547 & .000529 & .999986 \\
32,000 & .999545 & .999865 & .000131 & .999996 \\
\end{tabular}
\label{tabi}
\end{table}
\begin{table}
\caption[Table II]{Ionization and Unitarity: Probabilities for Ca + Ca}
\begin{tabular}{ccccc}
b(fm) & $e^-_{gr}$ & $\sum_{bnd} e^-$ & 
$\sum_{cont} e^-$ &
$\sum e^- $\\
250 & .95335 & .96291 & .03548 & .99839 \\
500 & .95351 & .96297 & .03516 & .99812 \\
1000 & .95657 & .96562 & .03278 & .99841 \\
2000 & .96355 & .97170 & .02707 & .99878 \\
4000 & .97583 & .98293 & .01635 & .99927 \\
8000 & .99028 & .99520 & .00462 & .99982 \\
16,000 & .99760 & .99924 & .00074 & .99998 \\
32,000 & .999419 & .999830 & .000165 & .999996 \\
64,000 & .9998559 & .9999585 & .0000404 & .9999989 \\
128,000 & .9999640 & .9999897 & .0000100 & .9999997 \\
\end{tabular}
\label{tabii}
\end{table}
\begin{table}
\caption[Table III]{Ionization and Unitarity: Probabilities for Ne + Ne}
\begin{tabular}{ccccc}
b(fm) & $e^-_{gr}$ & $\sum_{bnd} e^-$ & 
$\sum_{cont} e^-$ &
$\sum e^- $\\
500 & .98814 & .99058 & .00901 & .99959 \\
1000 & .98816 & .99058 & .00895 & .99952 \\
2000 & .98894 & .99125 & .00830 & .99960 \\
4000 & .99070 & .99278 & .00692 & .99970 \\
8000 & .99383 & .99564 & .00418 & .99982 \\
16,000 & .99753 & .99878 & .00117 & .99996 \\
32,000 & .999392 & .999809 & .000186 & .999995 \\
64,000 & .9998534 & .9999572 & .0000416 & .9999989 \\
128,000 & .9999636 & .9999895 & .0000102 & .9999997 \\
256,000 & .99999092 & .99999740 & .00000254 & .99999993 \\
\end{tabular}
\label{tabiii}
\end{table}
\begin{table}
\caption[Table IV]{Calculated Ionization Cross Sections Expressed in the
Form $A \ln \gamma + B$ (in barns)}
\begin{tabular}{|cc|ccccc|}
&& Pb + Pb & Zr + Zr & Ca + Ca & Ne + Ne & H + H\\
\tableline
&$1 - \sum_{bnd} e^-$ & 8680 & 10,240 & 10,620 & 10,730 & 10,770 \\
$\ \ A$&$\sum_{cont} e^-$& 8450 & 9970 & 10,340 & 10,440 & 10,480 \\
&Anholt \& Becker\cite{ab}& 13,800 & 11,600 & 10,800 & 10,600 & 10,540 \\
\tableline
&$1 - \sum_{bnd} e^-$& 14,190 & 28,450 & 38,010 & 46,080 & 71,090 \\
$\ \ B$&$\sum_{cont} e^-$& 12,920 & 27,110 & 36,530 & 44,430 & 68,780 \\
&Anholt \& Becker& 13,000 & 27,800 & 37,400 & 45,400 & 70,000\\
\end{tabular}
\label{tabiv}
\end{table}
\begin{table}
\caption[Table V]{Calculated values of the scaled quantity 
$(Z_2^2 / Z_1^2) B$ for non-symmetric combinations
of colliding particles.  The second nucleus ($Z_2$) is taken to be the one with
the single electron to be ionized.  
Since Anholt and Becker cross sections without screening are completely
perturbative, their values of of B also can be taken from Table IV, and are
repeated here for convenient comparison.}
\begin{tabular}{|c|cccccc|}
& H + Ne & H + Ca & Ca + H & H + Zr & H + Pb & Pb + H \\
\tableline
$1 - \sum_{bnd} e^-$ & 46,150 & 38,270 & 70,820 & 29,440 & 17,090 & 67,550\\
$\sum_{cont} e^-$& 44,490 & 36,790 & 68,520 & 28,070 & 15,680 & 65,330\\
Anholt \& Becker\cite{ab}& 45,400 & 37,400 & 70,000 & 27,800 & 13,000 & 70,000\\
\tableline
\tableline
& Pb +Ne & Ne + Pb & Pb + Ca & Ca + Pb & Pb + Zr & Zr + Pb \\
\tableline
$1 - \sum_{bnd} e^-$& 42,560 & 17,030 & 34,720 & 16,870 & 26,010 & 16,250\\
$\sum_{cont} e^-$& 41,000 & 15,690 & 33,330 & 15,530 & 24,730 & 14,930\\
Anholt \& Becker& 45,400 & 13,000 & 37,400 & 13,000 & 27,800 & 13,000\\
\end{tabular}
\label{tabv}
\end{table}
\begin{table}
\caption[Table VI]{Example: Calculated Ionization Cross Sections For Pb + Pb
(in barns)}
\begin{tabular}{cccc}
& AGS $\gamma = 11.3$& CERN $\gamma = 160$& RHIC $\gamma = 23,000$\\
$1 - \sum_{bnd} e^-$& 35,200 & 58,200 & 101,400 \\
$\sum_{cont} e^-$& 33,400 & 55,800 & 97,800 \\
Anholt \& Becker\cite{ab}& 46,700 & 83,700 & 151,600 \\
Anholt \& Becker (with screening)& & 63,600 &\\
Krause et al. Pb + Au  data\cite{kr}& & 42,000 &\\
\end{tabular}
\label{tabvi}
\end{table}
\end{document}